\documentclass[aps,prb,twocolumn,footinbib,showpacs,superscriptaddress,preprintnumbers,amsmath,amssymb]{revtex4-1}

\usepackage{graphicx}
\usepackage{subfigure}
\usepackage{dcolumn}
\usepackage{bm}
\usepackage{verbatim} 
\usepackage{color}

\begin{document}

\title{Thermodynamics of energy, charge and spin currents in thermoelectric quantum-dot spin valve}

\author{Gaomin Tang}
\affiliation{Department of Physics and the Center of Theoretical and Computational Physics, The University of Hong Kong, Hong Kong, China}
\author{Juzar Thingna}
 \email{juzar.thingna@uni.lu}
\affiliation{Complex Systems and Statistical Mechanics, Physics and Materials Science, University of Luxembourg, L-1511 Luxembourg, Luxembourg}
\author{Jian Wang}
 \email{jianwang@hku.hk}
\affiliation{Department of Physics and the Center of Theoretical and Computational Physics, The University of Hong Kong, Hong Kong, China}

\date{\today}

\begin{abstract}
We provide a thermodynamically consistent description of energy, charge and spin transfers in a thermoelectric quantum-dot spin valve in the collinear configuration based on nonequilibrium Green's function and full counting statistics. We use the fluctuation theorem symmetry and the concept of entropy production to characterize the efficiency with which thermal gradients can transduce charges or spins against their chemical potentials, arbitrary far from equilibrium. Close to equilibrium, we recover the Onsager reciprocal relations and the connection to linear response notions of performance such as the figure of merit. We also identify regimes where work extraction is more efficient far then close from equilibrium. 
\end{abstract}
\maketitle

\section{Introduction}
\label{sec:I}
Heat management and control in nano-scaled spin devices has become increasingly popular after the experimental discovery of spin Seebeck effect (SSE) \cite{SSE, SSE2} that led to the advent of spin caloritronics \cite{Bauer1, Bauer2}. The main motivation behind this field is to utilize waste heat in order to generate charge and spin current efficiently in magnetic nanostructures. The heat management ability of these devices finds novel applications in power converters \cite{Bauer3}, thermometers \cite{thermo_SC, thermo_SC1} and thermally assisted recording devices \cite{Bauer4, Bauer5, 9,10,11}. In order to make the harvesting process from heat as efficient as possible, nanostructured systems with reduced dimension such as quantum point contacts \cite{QPC1, QPC2} and quantum dots \cite{QD1, QD3, QD4, Zhou151, Zhou15} are particularly interesting due to their high figure of merit (FOM) $ZT$ and reduced phonon thermal conductivity \cite{phonon} in comparison with the bulk structures.

The dimensionless FOM $ZT$ is the most commonly used measure of performance of thermoelectric power generation. It is exclusively defined in terms of linear response coefficients and becomes inadequate far from equilibrium. The spin FOM $Z_sT$ was introduced to parameterize a thermally generated spin-current harvester in analogy with the FOM of a spinless system \cite{DiVentra, Barnas, Xiaobin1, Xiaobin2}. In nano devices as their size becomes comparable or smaller to the inelastic scattering length, deviations from the (linear response) Wiedeman-Franz law occur \cite{nonlinear1, nonlinear2} and nonlinear transport coefficients cannot be neglected anymore. It thus becomes essential to establish measures of performance in nonlinear thermoelectrics that remain meaningful beyond the linear response FOM. This can be done using concepts from thermodynamics of heat machines such as the efficiency and related quantities (maximum power, maximum efficiency, and minimum dissipation) \cite{QD3, MP1, MP2, MP3, MP4, ME1, ME2, PED, Benenti17, Colin1}.


Another effect arising at small scales is that the relative importance of fluctuations increase. ``Macroscopic'' efficiencies are defined as a ratio between the input and output ensemble averaged fluxes. While they provide proper measures of performance in the nonlinear regime, they are inadequate to characterize the fluctuations in the efficiency of a single device given by the ratio between an input and an output flux that both fluctuate. Full counting statistics (FCS) can be used to characterize current fluctuations \cite{Levitov2, Nazarov, Flindt1, RMP, Schmidt1, Schmidt2, Flindt2, Flindt3, Flindt4, gm2, yeyati_FCS, gm3, gm4, gm5} as well as efficiency fluctuations \cite{EF1, EF2, EF3, EF4, EF5, Bijay, EF6, time_COP}. The efficiency statistics provides a complete picture of the device performance and have been observed experimentally \cite{EF_exp, EF6}. The macroscopic efficiency is recovered as the most likely value of the efficiency statistics.

In this work, we provide a complete framework to study the statistics of charge and spin current efficiency for a quantum-dot spin valve (QDSV) in the collinear configuration. The efficiency fluctuations are studied from the perspective of FCS and we use the nonequilibrium Green's function (NEGF) formalism that does not require the week coupling of the system to the reservoirs \cite{Haug, Wang}. This allows us to capture the true quantum nature of efficiency statistics as compared to quantum master equations \cite{MassiJPCC, EF3} that are limited to the weak coupling regime. Furthermore, we obtain the scaled cumulant generating function (SCGF) of spin, charge and heat current that obeys the fluctuation theorem and thus allows us to obtain a thermodynamically consistent definition of charge and spin current efficiencies via the entropy production rate (EPR). For both efficiencies, we obtain the large deviation functions (LDF) from the SCGF that allows us to study the efficiency statistics. We numerically show that the device could be made more efficient beyond the linear response limit and analyze the charge and spin current efficiency fluctuations.

The paper is structured as follows. In Sec.~\ref{sec:II}, the model Hamiltonian and the SCGF for the particle, spin and heat current will be given in terms of NEGF and the fluctuation symmetry will be checked. Charge and spin current efficiency will be defined and their corresponding FOM will be obtained in the linear response regime in Sec.~\ref{sec:III}. Section~\ref{sec:IV} is devoted to the charge and spin current efficiency statistics. Numerical results are shown in Sec.~\ref{sec:V} together with their physical interpretations. We finally summarize our work in Sec.~\ref{sec:VI}.

\section{Model and transport}
\label{sec:II}

\subsection{The model}

We consider an insulating layer modeled as a quantum dot sandwiched between two ferromagnetic layers modeled as electrodes. The setup is in the collinear configuration, meaning that the magnetic moment of the left electrode is parallel or anti-parallel to the one of the right electrode. A tunneling magnetoresistance effect ensues because the resistance of the device depends on the relative orientation of the two magnetic moments \cite{TMR1,TMR2}. We assume that no spin flip occurs across the tunnel junction. The Hamiltonian of the whole system reads
\begin{equation}
\label{eq:1}
H=H_L+H_R+H_C+H_{T} .
\end{equation}
The Hamiltonian of the left and right electrode, $H_\alpha$ ($\alpha=L/R$), and of the central quantum dot, $H_C$, are given by
\begin{align}
\label{eq:2}
H_{\alpha} &= \sum_{k\sigma} \epsilon_{k\alpha\sigma} c_{k\alpha\sigma}^{\dag} c_{k\alpha\sigma} , \nonumber \\
H_C &= \sum_{\sigma} \epsilon_{\sigma} d_{\sigma}^{\dag} d_{\sigma} ,
\end{align}
where $\epsilon_\sigma =\epsilon +\frac{1}{2}\sigma\Delta\epsilon$ is the spin-dependent onsite energy with $\Delta \epsilon$ being the energy difference between the spin-up and spin-down electron energy levels inside the quantum dot. The splitting between the spin-up and spin-down energies can be induced via an external magnetic field or could be intrinsic in magnetic quantum dots. In turn, $H_T$ is the tunneling (without spin flip) Hamiltonian between the electrodes and the quantum dot 
\begin{equation}
\label{eq:3}
H_{T} =\sum_{k\alpha\sigma}(t_{k\alpha\sigma} c_{k\alpha\sigma}^{\dag} d_{\sigma} +H.c.).
\end{equation}
The electrodes are initially assumed at equilibrium at time $t=0$
\begin{equation}
\rho_{\alpha}=\exp{\left[- \frac{H_\alpha - \sum_{\sigma} \mu_{\alpha\sigma} N_{\alpha\sigma}}{T_{\alpha}}\right]}/Z,
\end{equation}
where $N_{\alpha\sigma}=\sum_{k} c_{k\alpha\sigma}^{\dag} c_{k\alpha\sigma}$ is the electron number in electrode $\alpha$ with spin $\sigma$ and $Z$ is the normalizing partition function. The chemical potential of electrode $\alpha$ in spin direction $\sigma$ is denoted as $\mu_{\alpha\sigma}$ and its temperature $T_{\alpha}$. Throughout the paper we set Boltzmann and Planck constant as well as electronic charge to unity: $k_B=\hbar=e=1$. In real materials, there will be an additional contribution to heat transport due to phonons that we have ignored herein and focused on the electronic contributions only.


\subsection{Counting statistics}

The statistics of the charge, spin and energy transfers is obtained from the differences in the outcomes between a projective measurement of $N_{L\sigma}$ and $H_L$ (which commute) at time $0$ and at time $t$. Because we will eventually only focus on steady state properties, measurements at the right electrode do not provide additional information \cite{PRE92}. The generating function (GF) of these transfer probability is given by \cite{RMP, gm2,gm3,gm4},
\begin{equation}
{\cal Z}(t)={\rm Tr}\left[\rho(0) \hat{U}_{\gamma}(0,t)\hat{U}_{\gamma}(t,0) \right].
\end{equation}   
The dressed evolution operator is given by 
\begin{align}
\label{eq:5}
\hat{U}_{\gamma}(t,t') &= e^{\sum_\sigma i\gamma_\sigma N_{L\sigma}+i\gamma_E H_L} \hat{U}(t,t') e^{-\sum_\sigma i\gamma_\sigma N_{L\sigma}-i\gamma_E H_L} \notag \\
&= {\mathbb T}_C \exp\left[-i\int_{t'}^t \hat{H}_\gamma(t_1)dt_1  \right],
\end{align}
where the dressed Hamiltonian reads
\begin{align}
\label{eq:6}
\hat{H}_{\gamma}(t) = e^{\sum_\sigma i\gamma_\sigma N_{L\sigma}+i\gamma_E H_L} \hat{H}(t) e^{-\sum_\sigma i\gamma_\sigma N_{L\sigma}-i\gamma_E H_L}
\end{align}
and ${\mathbb T}_C$ is the time-ordering operator.
The counting fields $\{ \gamma_\uparrow,\gamma_\downarrow,\gamma_E \}$ which ``dress'' the evolution keep track, respectively, of the number of spin-$\uparrow$ and spin-$\downarrow$ electrons as well as the energy entering the left electrode.
They depend on the Keldysh contour branch on which they reside and take the value $\{-\lambda_\uparrow/2, -\lambda_\downarrow/2, -\lambda_E/2 \}$ on the forward branch and $\{ \lambda_\uparrow/2, \lambda_\downarrow/2, \lambda_E/2 \}$ on the backward branch. 
The scaled cumulant generating function (SCGF) characterizing the large deviations of the stead-state currents can be expressed as
\begin{eqnarray}
\label{eq:7}
{\cal F}(\lambda_\uparrow,\lambda_\downarrow,\lambda_E) &=&\lim_{t\rightarrow \infty}\frac{\ln{\cal Z}(t)}{t} \nonumber \\
&=& \int \frac{d\omega}{2\pi} \ln \left[ \frac{\det(G^{-1}_\lambda)}{\det(G^{-1}_{\lambda=0})}  \right] ,
\end{eqnarray}
in terms of the dressed inverse Green's functions 
\begin{equation}
\label{eq:8}
G^{-1}_\lambda=\begin{pmatrix}
G^{a,-1} - \Sigma^> & \widetilde{\Sigma}_L^< + \Sigma_R^<   \\
\widetilde{\Sigma}_L^> + \Sigma_R^>  &  - G^{r,-1} - \Sigma^>
\end{pmatrix}.
\end{equation}
The dressed lesser and greater self-energy in spin space is given by,
\begin{align}
\label{eq:9}
\widetilde{\Sigma}_L^< &=
\begin{pmatrix}
e^{i\lambda_\uparrow + i\lambda_E\omega} \Sigma_{L\uparrow}^<  & 0  \\
0   &   e^{ i\lambda_\downarrow +i\lambda_E\omega} \Sigma_{L\downarrow}^<
\end{pmatrix} ,  \notag \\
\widetilde{\Sigma}_L^> &=
\begin{pmatrix}
e^{ -i\lambda_\uparrow - i\lambda_E\omega} \Sigma_{L\uparrow}^>  & 0  \\
0   &   e^{ -i\lambda_\downarrow - i\lambda_E\omega} \Sigma_{L\downarrow}^>
\end{pmatrix} .
\end{align}
We note that the four sub-block matrices of $G^{-1}_\lambda$ are symmetric in spin space. Since the system is collinear, there are no off-diagonal terms to the self-energy in the spin space implying that the spin-up and -down electrons are independent of each other. The undressed Green's functions and self-energies used in Eqs.~\eqref{eq:8} and \eqref{eq:9} can be expressed as,
\begin{align}
\label{eq:9.1}
G^r_{\sigma\sigma} = [G^a_{\sigma\sigma}]^* &= \frac{1}{\omega-\epsilon_\sigma+i(\Gamma_{L\sigma}+\Gamma_{R\sigma})/2},\notag \\
\Sigma_{\alpha\sigma}^< &= i \Gamma_{\alpha\sigma}f_{\alpha\sigma},\notag \\ 
\Sigma_{\alpha\sigma}^> &= i \Gamma_{\alpha\sigma}(f_{\alpha\sigma}-1),
\end{align}
where $\Gamma_{\alpha\sigma}$ is the coupling strength which we assume energy independent (wide-band approximation). Throughout we will suppress the explicit dependence of the frequency $\omega$ on the Green's function, fermi distributions, and the self-energies for notational simplicity. 
Using the above equations we can further simplify Eq.~\eqref{eq:7} as,
\begin{align}
\label{eq:12}
{\cal F}(\{\lambda_0 \}) = \int \frac{d\omega}{2\pi} \ln & \Big\{ 1+ \sum_\sigma \big[
(e^{ i\lambda_\sigma +i\omega\lambda_E}-1){\cal T}_{\sigma} f_{L\sigma}\bar{f}_{R\sigma}  \notag \\
+&(e^{-i\lambda_\sigma -i\omega\lambda_E}-1){\cal T}_{\sigma} f_{R\sigma}\bar{f}_{L\sigma}\big]\Big\},
\end{align}
with the short-hand notation $\bar{f}_{\alpha\sigma} = 1-f_{\alpha\sigma}$ for the Fermi-Dirac distribution in the $\alpha$-electrode and $\{\lambda_0 \} =(\lambda_\uparrow, \lambda_\downarrow, \lambda_E)$. The transmission coefficients (explicit $\omega$ dependence suppressed) are given by
\begin{align}
\label{eq:13}
{\cal T}_{\uparrow} &= \Gamma_{L\uparrow}\Gamma_{R\uparrow} |G^r_{\uparrow\uparrow}|^2 , \quad
{\cal T}_{\downarrow} =\Gamma_{L\downarrow}\Gamma_{R\downarrow} |G^r_{\downarrow\downarrow}|^2.
\end{align}

Using the identity,
\begin{equation}
\label{eq:10}
f_{L\sigma}(1-f_{R\sigma})=e^{\beta_L\mu_{L\sigma}-\beta_R\mu_{R\sigma}} f_{R\sigma}(1-f_{L\sigma}),
\end{equation}
we find the fluctuation theorem symmetry of the SCGF
\begin{align}
\label{eq:11}
& {\cal F}(\lambda_\sigma,\lambda_E) = \nonumber \\
& {\cal F}(-\lambda_\sigma+i(\beta_R\mu_{R\sigma}-\beta_L\mu_{L\sigma}),-\lambda_E+i(\beta_L-\beta_R) ),
\end{align}
with $\beta_\alpha = (k_B T_\alpha)^{-1}$. 
This symmetry implies that the backward joint probability of the transferred spin-up and -down electrons ($N_\sigma$) and energy ($E$) is exponentially disfavored with respect to the forward one with the relation
\begin{align}  \label{forward}
&\frac{P(N_\uparrow,N_\downarrow,E)}{P(-N_\uparrow,-N_\downarrow,-E)} =\exp{[\dot{\bf s}_i \; t]}
 \notag \\
=& \exp\left[\sum_\sigma(\beta_R\mu_{R\sigma}-\beta_L\mu_{L\sigma})N_\sigma + (\beta_L-\beta_R)E\right] ,
\end{align}
where $\dot{\bf s}_i$ is entropy production rate. The above relation, known as the steady-state fluctuation theorem, also tells us that the probability of an entropy production increase is exponentially more likely than the probability of entropy production decrease. A generalized version of the above fluctuation theorem in presence of a magnetic field has been derived from the principles of microscopic reversibility in Ref.~[\onlinecite{Chenjie}]. The second law of thermodynamics, i.e., in average the entropy production rate is non-negative $\langle \dot{\bf s}_i \rangle \equiv \dot{\bf S}_i \geq 0$, is a simple consequence of Jensen's inequality applied to the fluctuation theorem [Eq.~\eqref{forward}]. 

Partially differentiating the SCGF with respect to the counting field, i.e., $I_{\sigma}=\partial_{i\lambda_\sigma}{\cal F}|_{\lambda_\uparrow =\lambda_\downarrow =\lambda_E=0}$ and $I_{E}=\partial_{i\lambda_E}{\cal F}|_{\lambda_\uparrow =\lambda_\downarrow =\lambda_E=0}$, we obtain the spin-up and -down electronic current as,
\begin{align}
\label{eq:15}
I_{\uparrow} &= \int\frac{d\omega}{2\pi} {\cal T}_{\uparrow} (f_{L\uparrow}-f_{R\uparrow}) ,\nonumber \\
I_{\downarrow} &= \int\frac{d\omega}{2\pi} {\cal T}_{\downarrow}(f_{L\downarrow}-f_{R\downarrow}) ,
\end{align}
and the energy current
\begin{equation}
\label{eq:16}
I_{E} =\int\frac{d\omega}{2\pi} \hbar\omega \sum_\sigma {\cal T}_{\sigma} (f_{L\sigma}-f_{R\sigma}).
\end{equation}
The heat current is given by
\begin{equation}
I_h = I_E - \sum_\sigma\mu_{R\sigma} I_\sigma . \label{heat}
\end{equation}
In our definition above, $-I_h$ is the heat current taken from the hot electrode to fuel the device. This definition arises from the first law of thermodynamics, namely, energy conservation for the total system i.e., $I_E=I_h - \dot{W}_{chem}$ with the total rate of chemical work $\dot{W}_{chem} = -\sum_{\sigma}\mu_{R\sigma}I_{\sigma}$.

\subsection{Linear regime}

Equilibrium occurs when $T_R=T_L$ and $\mu_{L\sigma}=\mu_{R\sigma}$. Introducing $\Delta T = T_R-T_L$ and $\Delta \mu_\sigma = \mu_{L\sigma} -\mu_{R\sigma}$, we can expand the Fermi-Dirac distributions around $T=(T_L+T_R)/2$ and $\mu_{\sigma}=(\mu_{L\sigma}+\mu_{R\sigma})/2$,
\begin{equation}
\label{eq:17}
f_{L\sigma}-f_{R\sigma} \approx f' \left[(\omega-\mu_{\sigma}) \Delta T/T^2 - \Delta \mu_{\sigma}/T \right].
\end{equation}
This corresponds to the linear regime where the fluxes are proportional to the corresponding thermodynamic forces via the Onsager matrix $\mathbf{L}$ with the expression
\begin{equation}
\label{eq:18}
\begin{bmatrix}
&I_\uparrow  \\  &I_\downarrow  \\ &I_h
\end{bmatrix} =
\begin{bmatrix}
L_{\uparrow} & 0 & L_{\uparrow h}  \\
0 & L_{\downarrow} & L_{\downarrow h} \\
L_{\uparrow h} & L_{\downarrow h} & L_{hh}
\end{bmatrix}
\begin{bmatrix}
&\Delta\mu_\uparrow/T \\ &  \Delta\mu_\downarrow/T  \\  & -\Delta T/T^2
\end{bmatrix} ,
\end{equation}
where
\begin{align}
L_{\sigma} &= -\int \frac{d\omega}{2\pi} f' {\cal T}_{\sigma} , \nonumber \\ 
L_{\sigma h}&=-\int\frac{d\omega}{2\pi} (\omega-\mu_{R\sigma}) f'{\cal T}_{\sigma},\\
L_{hh} &= -\int\frac{d\omega}{2\pi} f' \left[(\omega-\mu_{R\uparrow})^2{\cal T}_{\uparrow} + (\omega-\mu_{R\downarrow})^2{\cal T}_{\downarrow} \right] .\nonumber
\end{align}

The Onsager coefficients are related to the famous transport coefficients \cite{Weymann1, Weymann2, Weymann3},
\begin{align}
G_{\sigma} &= \left.\frac{I_{\sigma}}{\Delta\mu_{\sigma}}\right|_{\substack{\Delta T = 0 \\ \Delta \mu_{\sigma'} = 0 \\ \{\sigma'\neq \sigma\}}} = \frac{L_{\sigma}}{T}\\
K &= \left.-\frac{I_{h}}{\Delta T}\right|_{I_{\uparrow, \downarrow} = 0} = \frac{1}{T^2}\frac{\mathrm{det} \,\mathbf{L}}{L_{\uparrow}L_{\downarrow}} \\
S_\sigma &= \left.\frac{\Delta \mu_\sigma}{\Delta T}\right|_{I_{\sigma} = 0} = \frac{1}{T}\frac{L_{\sigma h}}{L_{\sigma}}, 
\end{align}
where $G_{\sigma}$ is the spin-$\sigma$ conductance, $S_\sigma$ is the spin-$\sigma$ Seebeck coefficient, and $K$ is the thermal conductance.

Introducing the charge $I_p$ and spin $I_s$ current
\begin{equation}
\label{eq:21}
I_p = I_\uparrow + I_\downarrow ,\quad
I_s = I_\uparrow - I_\downarrow , 
\end{equation}
together with the heat current (\ref{heat}), we also find that 
\begin{equation}
\label{eq:23}
\begin{bmatrix}
&I_p  \\  &I_s  \\ &I_h
\end{bmatrix} =
\begin{bmatrix}
T G_p & T G_s & G_p S T^2  \\
T G_s & T G_p & P' G_p  S T^2 \\
G_p ST^2 & P' G_p  ST^2 & \kappa T^2
\end{bmatrix}
\begin{bmatrix}
&\Delta\mu_p/T \\ &  \Delta\mu_s/T  \\  & -\Delta T/T^2
\end{bmatrix} ,
\end{equation}
where we introduced the charge and spin biases 
\begin{align}
\label{eq:22}
\Delta\mu_p &= \frac{1}{2} ( \Delta\mu_\uparrow + \Delta\mu_\downarrow ) , \nonumber \\
\Delta\mu_s &= \frac{1}{2} ( \Delta\mu_\uparrow - \Delta\mu_\downarrow ) ,
\end{align}
and the coefficients
\begin{align}
\label{eq:24}
G_p &= G_{\uparrow} + G_{\downarrow}, \ \ G_s = G_{\uparrow} - G_{\downarrow}, \ \ \kappa = L_{hh} \notag \\
S & = \frac{G_{\uparrow}S_{\uparrow} + G_{\downarrow}S_{\downarrow}}{G_{\uparrow}+G_{\downarrow}}, \ \  P'= \frac{G_{\uparrow}S_{\uparrow} - G_{\downarrow}S_{\downarrow}}{G_{\uparrow}S_{\uparrow} + G_{\downarrow}S_{\downarrow}}.
\end{align}
Above we have expressed the coefficients in term of conventional coefficients \cite{SSE2, Bauer2, Bauer3}:
$G_p$ and $G_s$ are the charge and spin conductance, $S$ is a factor appearing in the charge and spin Seebeck coefficients [see Eqs.~\eqref{eq:34} and \eqref{eq:38}], and $P'$ is the polarization of $G_p S$.
We note that the Onsager reciprocity relation is satisfied for the two Onsager matrices, Eq.~\eqref{eq:18} and \eqref{eq:23}, due to the fluctuation theorem symmetry (\ref{eq:11}) \cite{RMP}.
\section{Energy transduction}
\label{sec:III}

The entropy production rate (EPR), $\dot{\bf S}_i$, for a system at steady state between multiple reservoirs is given by the rate of entropy changes in those reservoir. For the QDSV in collinear configuration it is thus given by
\begin{equation}
\label{eq:28}
\dot{\bf S}_i= \sum_\alpha\beta_\alpha I_{\alpha h} \geq 0 ,
\end{equation}
where $I_{\alpha h}=I_{\alpha E}-\mu_{\alpha\uparrow} I_{\alpha\uparrow}-\mu_{\alpha\downarrow} I_{\alpha\downarrow}$. The non-negativity of the EPR follows from the fluctuation theorem symmetry, as discussed below Eq.~\eqref{forward}.

The EPR can be simplified using conservation laws that dictate the number of independent affinity and current pairs that contribute to EPR \cite{Matteo1}. In our model we have 6 affinities $\{\beta_{\alpha}\mu_{\alpha\uparrow}, \beta_{\alpha}\mu_{\alpha\downarrow}, \beta_{\alpha}\}$ for $\alpha = L,R$ and three conservation laws, namely, particle current conservation $I_{L(p)}+I_{R(p)} = 0$, spin current conservation $I_{L(s)}+I_{R(s)} = 0$, and energy current conservation $I_{L(h)}+I_{R(h)} = 0$, that give us 3 [6 (total affinities) - 3 (conservation laws)] independent affinity and current pairs. The EPR can thus be written as
\begin{align}
\label{eq:29}
& \dot{\bf S}_i = \dot{\bf S}_h + \dot{\bf S}_p + \dot{\bf S}_s ,\\
& \dot{\bf S}_h = -(\beta_L-\beta_R) I_h \\
& \dot{\bf S}_{p(s)} = \beta_L I_{p(s)} \Delta\mu_{p(s)} . 
\end{align}

When the QDSV operates as a charge (resp. spin) pump, $\dot{\bf S}_p < 0$ (resp. $\dot{\bf S}_s < 0 $). These processes can occur against their spontaneous direction because the remaining EPR contributions are sufficiently positive to make the total EPR positive. The macroscopic efficiency with which these processes occur is thus given by the ratio between the average output and input flow
\begin{align}
\label{eq:31}
0 \leq \bar{\eta}_p &= \frac{- \dot{\bf S}_p }{ \dot{\bf S}_h + \dot{\bf S}_s } 
=\frac{-\langle \dot{w}_p\rangle}{\eta_C I_h +\langle \dot{w}_s\rangle} \leq 1, \\ 
\label{eq:32}
0 \leq \bar{\eta}_s &= \frac{- \dot{\bf S}_s }{ \dot{\bf S}_h + \dot{\bf S}_p }
=\frac{- \langle \dot{w}_s\rangle}{\eta_C I_h +\langle \dot{w}_p\rangle} \leq 1,
\end{align}
where the the charge and spin power are given by 
\begin{align}
\label{eq:41}
\langle \dot{w}_p\rangle &= -I_p\Delta\mu_p, \quad \langle \dot{w}_s\rangle = -I_s\Delta\mu_s
\end{align}
and $\eta_C = (1-\beta_R/\beta_L)$ is the Carnot efficiency.
The upper bound one is a direct consequence of the non-negativity of the EPR.
These efficiencies are valid measure of performance arbitrary far from equilibrium and are not restricted to the linear response regime. 

Next, we obtain explicit expressions for the maximum charge and spin current efficiency in the linear response regime in terms of the FOM for charge $Z_p T$ and spin $Z_s T$. In order to obtain the charge FOM $Z_p T$, the spin current of the system should be zero. Using Eq.~\eqref{eq:23}, the condition $I_s=0$ can be satisfied if the spin bias
\begin{equation*}
\Delta\mu_s = P'S\Delta T-P\Delta\mu_p
\end{equation*}
where $P=G_s/G_p$ is the polarization of the spin conductance. Inserting the above expression into the charge current equation obtained via the Onsager matrix [Eq.~\eqref{eq:23}] we obtain,
\begin{align}
\label{eq:33}
I_p &= (1-P^2)G_p\Delta\mu_p - (1-PP')G_p S \Delta T  \notag \\
    &= G_{\rm spin}\Delta\mu_p - L_{Tp} \Delta T .
\end{align}
Thus, we can now define the charge Seebeck coefficient \cite{Barnas} as the ratio of the temperature bias coefficient to the chemical-potential bias coefficient when the charge current is zero, i.e., from Eq.~\eqref{eq:33} as, 
\begin{equation}
\label{eq:34}
S_p = \left.\frac{\Delta \mu_p}{\Delta T}\right|_{I_{p} = 0}=\frac{L_{Tp}}{G_{\rm spin}} = \frac{1-PP'}{1-P^2} S ,
\end{equation}
where we should note that $G_{\rm spin}=(1-P^2)G_p$ differs from $G_s$. Then the charge FOM $Z_pT$ can be defined as \cite{Bauer3},
\begin{align}
\label{eq:35}
Z_p T &= \frac{G_{\rm spin} S_p^2 T}{K}  \notag \\
&= \frac{(1-PP')^2 G_p S^2 T}{(2PP'-P'P'-1)G_p S^2 T + (1-P^2)\kappa}  .
\end{align}
In order to find the maximum charge current efficiency, we look for the zero of the derivative of $\bar{\eta}_p$ with respect to the charge current with zero spin current $I_s=0$, i.e., $\partial_{I_p} \bar{\eta}_p|_{I_s=0}=0$. Under this condition, we obtain the expression for the charge current,
\begin{equation}
\label{eq:36}
I_p =\frac{(1-PP') G_p S \Delta T}{Z_pT} (\sqrt{1+Z_pT}-1).
\end{equation}
Substituting the above expression into the definition of charge current efficiency [Eq.~\eqref{eq:31}], the maximum charge current efficiency (scaled by the Carnot) in the linear response regime reads \cite{Colin2, Colin3}
\begin{equation}
\label{eq:37}
(\bar{\eta}_p)_{\rm max} =\frac{\sqrt{1+Z_p T}-1}{\sqrt{1+Z_p T}+1} .
\end{equation}

Similar to the charge FOM and maximum charge current efficiency derived above we can also obtain the spin FOM and maximum spin current efficiency. In order to obtain these quantities we set the charge current $I_p=0$ to obtain a condition on the charge bias $\Delta \mu_p$. The condition then helps to obtain the spin-Seebeck coefficient when plugged into the expression for the spin current $I_s =0$ as,
\begin{equation}
\label{eq:38}
S_s = \left.\frac{\Delta \mu_s}{\Delta T}\right|_{I_{s} = 0} =\frac{P'-P}{1-P^2} S .
\end{equation}
Given the spin-Seebeck the spin FOM $Z_s T$ emerges naturally as,
\begin{align}
Z_s T &= \frac{G_{\rm spin} S_s^2 T}{K}  \notag \\
\label{eq:39}
&= \frac{(P'-P)^2 G_p S^2 T}{(2PP'-P'P'-1)G_p S^2 T + (1-P^2)\kappa} ,
\end{align}
which was also earlier proposed (without derivation) in Refs.~[\onlinecite{DiVentra}] and [\onlinecite{Barnas}]. Therefore, we can now easily obtain the maximum spin-current efficiency by setting the derivative of $\bar{\eta}_s$ with respect to the charge current to zero under the condition $I_p=0$, i.e., $\partial_{I_s} \bar{\eta}_s|_{I_p=0}=0$. This restricts the spin current $I_s=(P'-P)G_p S \Delta T (\sqrt{1+Z_sT}-1)/(Z_sT) ,$ and allows us to obtain the maximum spin-current efficiency (scaled by the Carnot) as, \cite{Colin2, Colin3}
\begin{equation}
\label{eq:40}
(\bar{\eta}_s)_{\rm max} =\frac{\sqrt{1+Z_s T}-1}{\sqrt{1+Z_s T}+1} .
\end{equation}
The above maximum spin-current efficiency has the same form as the charge-current efficiency with $Z_pT$ replaced by $Z_s T$. Thus, even in the spin case an increase in spin FOM $Z_sT$ ensures a high efficiency of the device in the linear response regime. We can also see from our expressions Eqs.~\eqref{eq:37} and \eqref{eq:40} that as $ZT \rightarrow \infty$ we approach Carnot efficiency.

Besides the maximum efficiency, another quantity of interest is the efficiency at maximum output power \cite{PED}. In the linear response regime, setting the derivative $\partial_{\Delta\mu_p} \langle \dot{w}_p\rangle  =0$ using Eq.~\eqref{eq:41} gives us the condition on the charge bias to achieve maximum output power, i.e., $\Delta\mu_p = \frac{1}{2}(S\Delta T -P\Delta\mu_s)$. Thus, the maximum charge power reads $\langle \dot{w}_p\rangle_{\rm MP} = \frac{1}{4}G_p (S\Delta T-P\Delta\mu_s)^2$. The corresponding charge current efficiency at maximum charge power is given by,
\begin{equation}
\label{eq:42}
(\bar{\eta}_p^l)_{\rm MP} = \frac{-\langle \dot{w}_p\rangle_{\rm MP}}{2\langle \dot{w}_p\rangle_{\rm MP} -\eta_C(\kappa\Delta T -2G_p P' S T\Delta\mu_s)-G_p\Delta\mu_s^2} . 
\end{equation}
Similarly, letting the derivative $\partial_{\Delta\mu_s} \langle \dot{w}_s\rangle =0$ gives us that the maximum output charge power happens at spin bias $\Delta\mu_s = \frac{1}{2}(P' S\Delta T -P\Delta\mu_p)$. Hence, the maximum spin power is $\langle \dot{w}_s\rangle_{\rm MP} = \frac{1}{4}G_p (P' S\Delta T-P\Delta\mu_p)^2$ and the corresponding spin current efficiency at maximum spin power is given by,
\begin{equation}
\label{eq:43}
(\bar{\eta}_s^l)_{\rm MP} = \frac{-\langle \dot{w}_s\rangle_{\rm MP}}{2\langle \dot{w}_s\rangle_{\rm MP} -\eta_C(\kappa\Delta T -2G_p S T\Delta\mu_p)-G_p\Delta\mu_p^2} . 
\end{equation}

In this section we defined the spin and charge efficiencies for the QDSV in linear response and far from equilibrium regimes. We explicitly showed that in the linear response regime the spin current efficiency is related to the spin FOM in an analogous way to the traditional methods relating the FOM $ZT$ and maximum thermoelectric efficiency. Moreover, we also found explicit expressions for the efficiencies at maximum power in the linear response regime that could again be expressed in terms of the Onsager coefficients. 

\section{Efficiency statistics}
\label{sec:IV}
In this section, we introduce the charge and spin current efficiency statistics from the perspective of FCS. We begin by  making the following counting field substitutions
\begin{align}
\label{eq:44}
i\lambda_\uparrow &\rightarrow -\lambda_h\mu_{R\uparrow}-\lambda_p\Delta\mu_p -\lambda_s\Delta\mu_s , \notag \\
i\lambda_\downarrow &\rightarrow -\lambda_h\mu_{R\downarrow}-\lambda_p\Delta\mu_p +\lambda_s\Delta\mu_s ,\notag \\
i\lambda_E &\rightarrow \lambda_h ,
\end{align}
in Eq.~\eqref{eq:12} to obtain the SCGF for charge, spin, and heat. Above, $\lambda_p,\lambda_s,\lambda_h$ are the counting fields for charge work $w_p = t \dot{w}_p$, spin work $w_s = t\dot{w}_s$, and heat $q=t\dot{q}$ and we set the flow direction from the right to the left reservoir as positive. The charge power $\dot{w}_p$, spin power $\dot{w}_s$, and heat current $\dot{q}$ are stochastic variables and their mean values are $|I_p\Delta\mu_p|$, $|I_s\Delta\mu_s|$ and $|I_h|$ that have been introduced in the last section. The SCGF after the substitutions has the form,
\begin{align}
\label{eq:45}
{\cal F}(\lambda_p,\lambda_s, &\lambda_h) = \notag \\
\int \frac{d\omega}{2\pi} \ln \Big[ 1 &+
(e^{-(\omega-\mu_{R\uparrow})\lambda_h +\lambda_p\Delta\mu_p+\lambda_s\Delta\mu_s}-1) {\cal T}_{\uparrow} f_{L\uparrow}\bar{f}_{R\uparrow} \notag \\
&+ (e^{(\omega-\mu_{R\uparrow})\lambda_h -\lambda_p\Delta\mu_p -\lambda_s\Delta\mu_s}-1) {\cal T}_{\uparrow} f_{R\uparrow}\bar{f}_{L\uparrow} \notag \\
&+ (e^{-(\omega-\mu_{R\downarrow})\lambda_h +\lambda_p\Delta\mu_p-\lambda_s\Delta\mu_s}-1) {\cal T}_{\downarrow} f_{L\downarrow}\bar{f}_{R\downarrow} \notag \\
&+ (e^{(\omega-\mu_{R\downarrow})\lambda_h -\lambda_p\Delta\mu_p +\lambda_s\Delta\mu_s}-1) {\cal T}_{\downarrow}f_{R\downarrow}\bar{f}_{L\downarrow}\Big] .
\end{align}
From Eq.~\eqref{eq:11}, the fluctuation theorem symmetry relation of SCGF can be expressed as,
\begin{equation}
\label{eq:46}
{\cal F}(\lambda_p,\lambda_s,\lambda_h) = {\cal F}(-\lambda_p+\beta_L,-\lambda_s+\beta_L,-\lambda_h+(\beta_L-\beta_R)) .
\end{equation}
Setting $\lambda_p=\lambda_s=\lambda_h=0$ this symmetry relation implies the integral fluctuation theorem, namely,
\begin{equation}
\label{eq:47}
\langle \exp\left[ -\beta_L (w_p + w_s) -(\beta_L-\beta_R) q \right]\rangle =1.
\end{equation}
Furthermore, the above equality helps bound the macroscopic value of efficiencies ($\bar{\eta}_{p(s)} = -\langle w_{p(s)} \rangle /(\eta_C\langle q \rangle + \langle w_{s(p)} \rangle)\leq 1$) defined in Eqs.~\eqref{eq:31} using the Jensen's inequality that implies $-(\langle w_p\rangle + \langle w_s \rangle)/(\eta_C \langle q \rangle) \leq 1$. In the probabilistic sense, since the fluctuation theorem relates the forward and backward joint probabilities (even though the backward is exponentially less likely) we expect to observe efficiencies lower and higher (exponentially unlikely) than the most likely ones in the statistics.

Since $w_p$, $w_s$, and $q$ are stochastic variables that fluctuate, so are the efficiencies $\eta_p = -w_p /(\eta_C q + w_s)$, $\eta_s = -w_s /(\eta_C q + w_p)$. Efficiency statistics are not bounded and can be characterized by the rate $J(\eta)$ at which the probability to observe a given efficiency $\eta$ decays exponentially during a long measurement time \cite{EF1, EF2, EF3},
\begin{equation}
\label{eq:48}
P(\eta) =\lim_{t\rightarrow \infty} \exp\left[-J(\eta)t\right] .
\end{equation}
This rate is called the large deviation function (LDF) of efficiency $\eta$ and can be related to the SCGF in Eq.~\eqref{eq:45}. Large deviation principle describes the exponentially unlikely deviations of a stochastic variable from its most likely value at which the large deviation function vanishes. We first show how we can get the LDF of charge efficiency $J(\eta_p)$ by setting the constraints on $\lambda_h$ and $\lambda_s$. The probability distribution of the charge efficiency in the long time limit is given by
\begin{equation}
P(\eta_p) = \iiint d\omega_p d\omega_s dq P(\omega_p, \omega_s, q) \delta \left(\eta_p - \frac{-\omega_p}{\eta_C q +\omega_s} \right), 
\end{equation} 
where the joint probability for charge-, spin-work and heat
\begin{align}
&P(\omega_p, \omega_s, q) =  \notag \\
&\iiint d\lambda_p d\lambda_s d\lambda_h e^{-\lambda_p\omega_p-\lambda_s\omega_s-\lambda_h q} e^{-t{\cal F}(\lambda_p, \lambda_s, \lambda_h)},
\end{align}
with the SCGF ${\cal F}(\lambda_p, \lambda_s, \lambda_h)$ given by Eq.~\ref{eq:45}. Thus, the probability distribution and the generating function are related simply by the Laplace transform, i.e.,
\begin{equation}
P(\eta_p) = \int d\lambda_p \exp[ -t{\cal F}(\lambda_p, \eta_p\lambda_p, \eta_p\eta_C\lambda_p) ].
\end{equation}
In the long time limit, using the Laplace approximation the integral simplifies to,
\begin{equation}
P(\eta_p) \asymp \lim_{t \rightarrow \infty} \exp[-t \ \mathrm{min}_{\lambda_p}{\cal F}(\lambda_p, \eta_p\lambda_p, \eta_p\eta_C\lambda_p)].
\end{equation} Then the LDF of charge efficiency $\eta_p$ is obtained through ${\cal F}(\lambda_p,\lambda_s,\lambda_h)$ by setting $\lambda_h=\eta_p\eta_C\lambda_p$, $\lambda_s=\eta_p\lambda_p$, and minimizing ${\cal F}$ relative to $\lambda_p$, namely \cite{EF1, EF2, EF3},
\begin{equation}
\label{eq:49}
J(\eta_p)=-\mathrm{min}_{\lambda_p}{\cal F}\left(\lambda_p, \eta_p\lambda_p, \eta_p\eta_C\lambda_p \right) .
\end{equation}
The minimization procedure requires us to set $d_{\lambda_p} {\cal F}\left(\lambda_p, \eta_p\lambda_p, \eta_p\eta_C\lambda_p \right)=0$ and obtain the constraint on $\lambda_p$ that minimizes ${\cal F}\left(\lambda_p, \eta_p\lambda_p, \eta_p\eta_C\lambda_p \right)$. The LDF for spin current efficiency is obtained in a similar way,
\begin{equation}
\label{eq:50}
J(\eta_s)=-\mathrm{min}_{\lambda_s}{\cal F}\left(\eta_s\lambda_s, \lambda_s, \eta_s\eta_C\lambda_s \right) .
\end{equation}
The efficiency at which we obtain the maximum (minimum) of the LDF is the Carnot efficiency $\eta_C$ (macroscopic efficiency), and this universal result follows directly from the fluctuation theorem \cite{EF1}. Since we want to get the minimum of ${\cal F}$ with respect to $\lambda_{p}$ in Eq.~\eqref{eq:49}, we first numerically obtain the value of $\lambda_{p}\in (-\beta_L/\eta_p, +\beta_L/\eta_p)$ satisfying the condition $d_{\lambda_p}{\cal F}(\lambda_p,\eta_p\lambda_p,\eta_p\eta_C\lambda_p)=0$, using the bisection method. Then we use the obtained $\lambda_{p}$ to get the minimum of ${\cal F}$ in Eq.~\eqref{eq:49}, and hence the LDF. The spin efficiency LDF in Eq.~\eqref{eq:50} is obtained in a similar manner.

In the linear response regime, by expanding the SCGF [Eq.~\eqref{eq:45}] we obtain
\begin{equation}
\label{eq:51}
{\cal F}(\lambda_p,\lambda_s, \lambda_h) = \frac{1}{2}\lambda {\bf C}\lambda^T +\lambda {\bf x}^T ,
\end{equation} 
with $\lambda=[\lambda_p,\lambda_s, \lambda_h]$. The real symmetric covariance matrix 
\begin{equation}
\label{eq:52}
{\bf C} =\begin{bmatrix}C_{pp} & C_{ps} & C_{ph}  \\ C_{sp} & C_{ss} & C_{sh}  \\  C_{hp} & C_{hs} & C_{hh}\end{bmatrix}
\end{equation}
and current vector ${\bf x}=[\langle \dot{w}_p\rangle, \langle \dot{w}_s\rangle, I_h]$. The covariance matrix entries can be obtained by relating the equilibrium fluctuations to the linear response coefficients through the Green-Kubo relation \cite{EF1}. Hence, the entries in terms of the Onsager matrix elements Eq.~\eqref{eq:23} are given by,
\begin{align}
\label{eq:53}
C_{pp}&=2G_pT\Delta\mu_p^2, \ \ &C_{ps}&=2G_sT\Delta\mu_p\Delta\mu_s, \notag \\
C_{ss}&=2G_pT\Delta\mu_s^2,\ \ &C_{sh}&=2G_p P' S T^2\Delta\mu_s, \notag \\
C_{hh}&=2\kappa T^2, \ \ &C_{ph}&=2G_p S T^2\Delta\mu_p.
\end{align}
Thus, in terms of the covariance matrix elements the SCGF,
\begin{align}
\label{eq:54}
{\cal F}\left(\lambda_p, \eta_p\lambda_p,\eta_p\eta_C\lambda_p \right)&=a_p\lambda_p^2+b_p\lambda_p, \notag \\
{\cal F}\left(\eta_s\lambda_s, \lambda_s,\eta_s\eta_C\lambda_s \right)&=a_s\lambda_s^2+b_s\lambda_s, 
\end{align}
with 
\begin{align}
\label{eq:55}
a_p =& \frac{1}{2}[C_{pp} +\eta_p^2 C_{ss} + \eta_p^2\eta_C^2 C_{hh}]  \notag \\
&+\eta_p[C_{ps}-\eta_C C_{ph}- \eta_p\eta_C C_{sh} ] ,  \notag \\
b_p =& \langle \dot{w}_p\rangle +\eta_p \langle \dot{w}_s\rangle +\eta_p\eta_C I_h , \notag \\
a_s =& \frac{1}{2}[C_{ss} +\eta_s^2 C_{pp} + \eta_s^2\eta_C^2 C_{hh}]  \notag \\
&+\eta_s[C_{ps}-\eta_C C_{sh}- \eta_s\eta_C C_{ph} ] ,  \notag \\
b_s =& \langle \dot{w}_s\rangle +\eta_s \langle \dot{w}_p\rangle +\eta_s\eta_C I_h .
\end{align}
From Eq.~\eqref{eq:49} and Eq.~\eqref{eq:50} the efficiency statistics in the linear response regime reads
\begin{equation}
\label{eq:56}
J(\eta_p)=\frac{b_p^2}{4a_p}, \qquad J(\eta_s)=\frac{b_s^2}{4a_s}.
\end{equation}
In the linear response limit, the entropy production can be approximated as $\dot{\bf S}_i = -\beta(\eta_C I_h + \langle \dot{w}_p\rangle +\langle \dot{w}_s\rangle )$ with $\eta_C = \Delta T/T$. Plugging Eqs.~\eqref{eq:23}, \eqref{eq:41}, \eqref{eq:53}, and \eqref{eq:55} into Eq.~\eqref{eq:56}, one can directly verify that both efficiency LDFs, $J(\eta_{p(s)})$, take the upper-bound value $\dot{\bf S}_i / 4$ at the Carnot $\eta_{p(s)} = 1$. Similar behavior was reported in the traditional thermoelectric devices without spin degree of freedom in the linear response limit\cite{Bijay}.

\section{Numerical Results}
\label{sec:V}
In this section, we test our analytic results numerically for the QDSV model described via Eqs.\eqref{eq:1}, \eqref{eq:2}, and \eqref{eq:3}. We restrict the spin-dependent bias to the left electrode, i.e., $\mu_{R\uparrow}=\mu_{R\downarrow}=\mu_R=0$, such that $\Delta\mu_p=(\mu_{L\uparrow}+\mu_{L\downarrow})/2$ and $\Delta\mu_s=(\mu_{L\uparrow}-\mu_{L\downarrow})/2$. The linewidth amplitude $\Gamma_{L\uparrow}=\Gamma_{L\downarrow}=\Gamma_0$, $\Gamma_{R\uparrow} = \Gamma_0 (1+p)$, and $\Gamma_{R\downarrow} = \Gamma_0 (1-p)$ with $p\in [-1,1]$ denoting the spin polarization degree of the right ferromagnetic electrode. Throughout we set the charge $\Delta\mu_p$ and spin bias $\Delta\mu_s$ to be positive, i.e., currents flow from left to right, opposite in direction to the temperature bias $\Delta T$. Thus, an output spin or charge power less than zero implies that the engine heats the hot and cold electrodes utilizing the spin or charge work. Such a system can no longer operate as a thermoelectric engine and is termed as a dud engine. The spin Seebeck coefficient could be up to $3.4meV/K$ in spin-semiconducting graphene nanoribbons \cite{Xiaobin2}, so that several $meV$ charge and spin bias as shown below is experimentally achievable.

\begin{figure}[tb!]
  \includegraphics[width=\columnwidth]{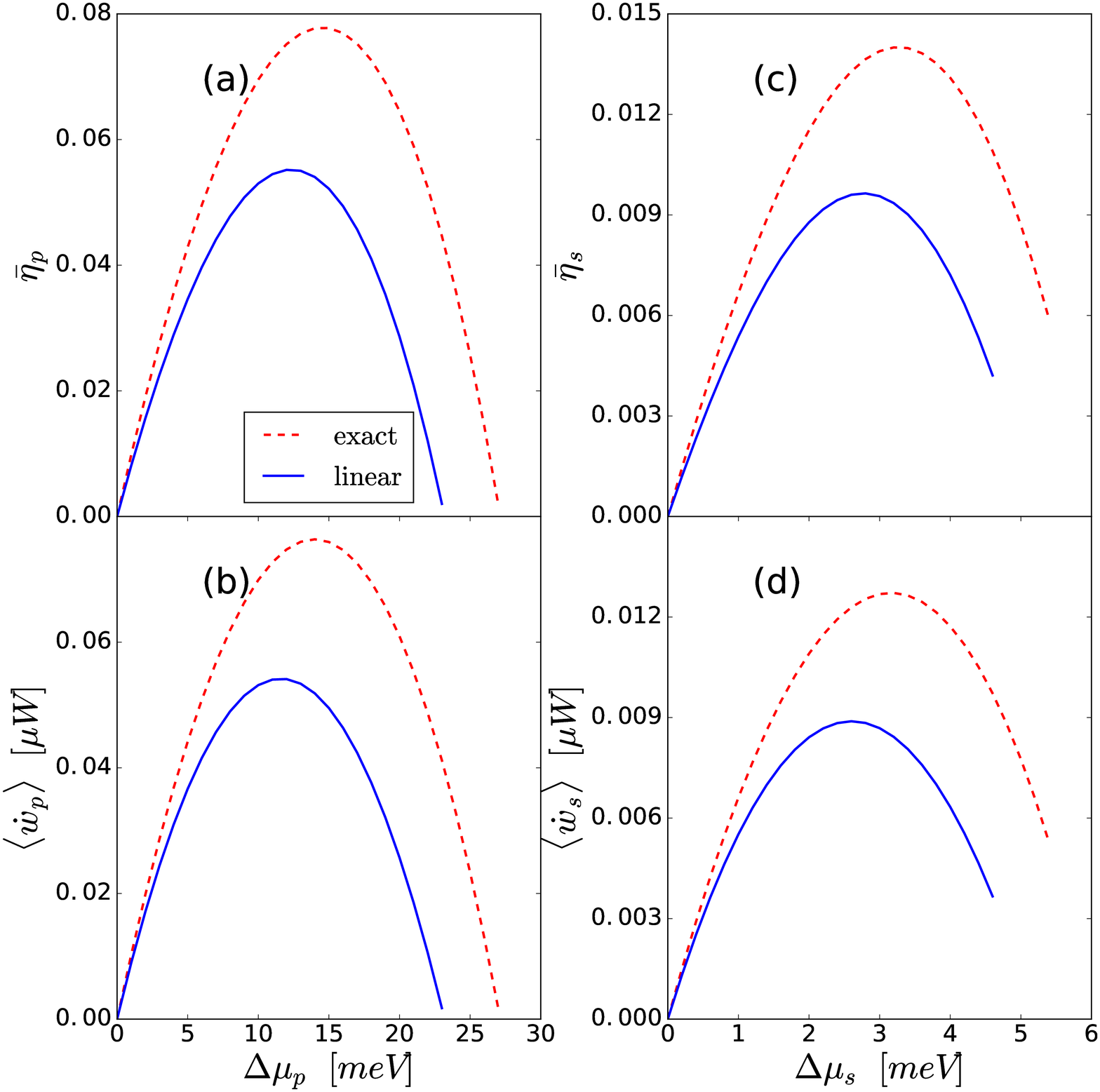}
  \caption{(Color online) Charge efficiency $\bar{\eta}_p$ [panel (a)], charge power $\langle \dot{w}_p\rangle$ [panel (b)], spin efficiency $\bar{\eta}_s$ [panel (c)], and spin power $\langle \dot{w}_s\rangle$ [panel (d)] are plotted versus the charge or spin bias using the exact formula (red dashed lines) and linear response limit (blue solid lines). The spin bias is chosen to be $\Delta\mu_s=0.2\Delta\mu_p$. The other parameters are $\epsilon_\uparrow=0.3eV$, $\epsilon_\downarrow=0.4eV$, $p=0.2$, $\Gamma_0=10meV$, $T_L=300K$, and $T_R=500K$.}
    \label{linear}
\end{figure}
Fig.~\ref{linear} compares the efficiencies ($\bar{\eta}_p$ and $\bar{\eta}_s$) and average power ($\langle\dot{w}_p\rangle$ and $\langle\dot{w}_s\rangle$) within the linear response approximation (near equilibrium) with the exact results. The maximum efficiency and power lie beyond the linear response implying an efficient engine far from equilibrium. Additionally, the linear response always tends to underestimate the power and efficiency. The underestimation of power is due to the positive higher order contributions of the spin and charge currents beyond linear response. This is naturally expected since the model doesn't exhibit any exotic features like negative differential resistance that requires a negative contribution from the higher order terms. The underestimation of efficiency is mainly due to the output power ($\dot{w}_{p/s}$) since the input flow of both spins and particles ($ \dot{\bf S}_h + \dot{\bf S}_{s/p} =\eta_C I_h +\langle \dot{w}_{s/p}\rangle$) does not show a systematic difference between the exact and linear response.

\begin{figure}[tb!]
  \includegraphics[width=\columnwidth]{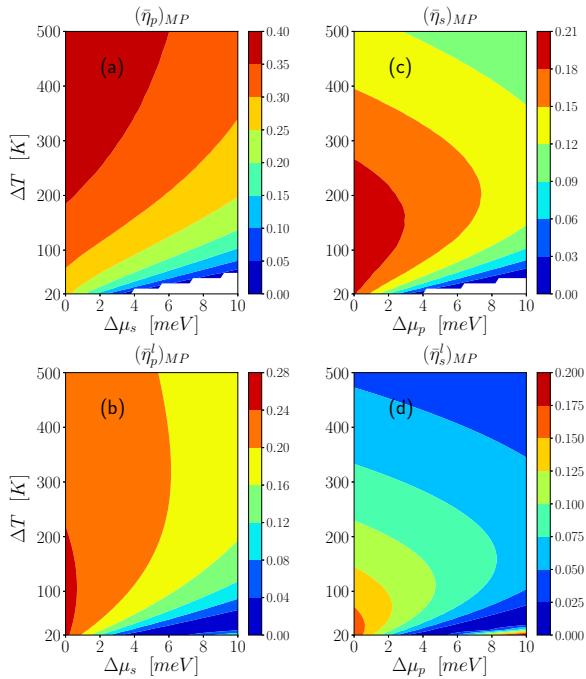}
  \caption{(Color online) Charge (spin) efficiency at maximum power $(\bar{\eta}_{p(s)})_{\rm MP}$ [panels (a) and (c)] and (b) their linear response limit $(\bar{\eta}_{p(s)}^l)_{\rm MP}$ [panels (b) and (d)] plotted against different spin (charge) $\Delta\mu_{s(p)}$ and temperature $\Delta T$ biases. The temperature in the left electrode is fixed to be $T_L=300K$. The other parameters are $\epsilon_\uparrow=0.1eV$, $\epsilon_\downarrow=0.2eV$, $p=0.2$, $\Gamma_0=10meV$.}
    \label{max}
\end{figure}

Since the maximal power and maximal efficiencies appear at different biases we plot the efficiency at maximum power as seen in Fig.~\ref{max}. From a practical standpoint, having maximum power output is the quantity of prime interest since it provides a reasonable measure of how efficient the device is when it is producing maximum power output. The linear response results again underestimate the efficiency at maximum power, implying that the exact theory provides valuable insight into the device performance. The linear response predicts that the thermoelectric and thermospin devices have an optimal working temperature bias beyond which the device degrades due to a substantial increase in heat contribution $\eta_C\kappa\Delta T$ [see Eqs.~\eqref{eq:42} and \eqref{eq:43}]. This feature persists far from equilibrium [Fig.~\ref{max}(a) and (c)] for the thermospin device but not for the thermoelectric one that monotonously becomes more efficient with the increase in temperature bias. The un-plotted regions in Figs.~\ref{linear} [beyond $27meV$ for $\Delta\mu_p$ in panels (a) and (b), and beyond $5.4meV$ for $\Delta\mu_s$ in panels (c) and (d)] and \ref{max} [white region in panels (a) and (c)] correspond to the the dud engine regime. Here, the large charge (spin) bias leads to a large power causing the system to be driven by the charge (spin) bias instead of the thermal bias. This causes the electrodes to heat up leading to an unphysical dud engine scenario. The linear response theory [Fig.~\ref{max}(b) and (d)] is unable to predict this breakdown of the engine and moreover estimates wrongly that the efficiency at maximum power for charge $(\eta^l_p)_{\rm MP}$ [spin $(\eta^l_s)_{\rm MP}$] increases with $\Delta\mu_p$ [$\Delta\mu_s$].

\begin{figure}[tb!]
  \includegraphics[width=\columnwidth]{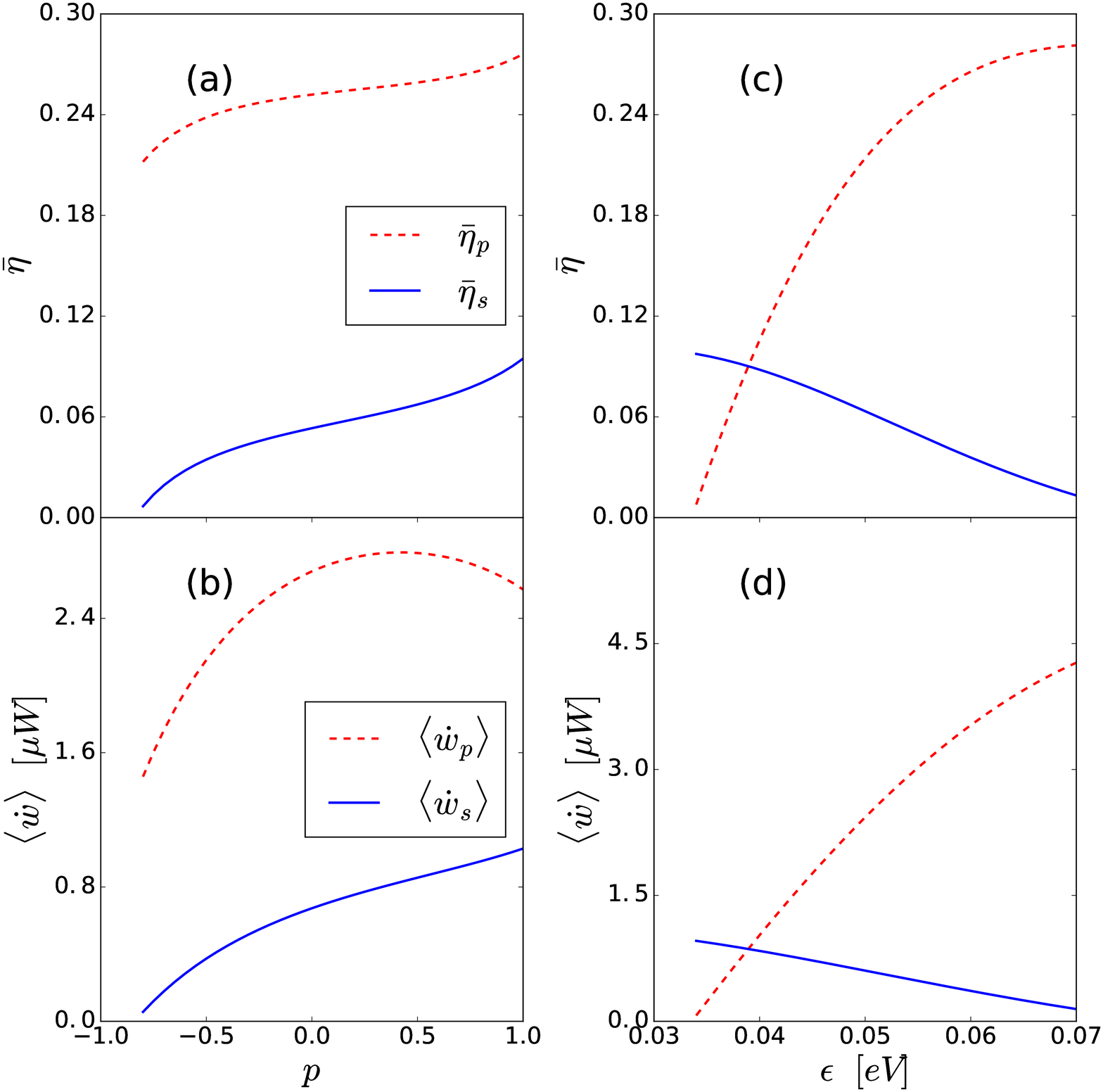}
  \caption{(Color online) Charge and spin efficiency $\bar{\eta}_{p/s}$ [panel (a)] and power $\langle \dot{w}_{p/s}\rangle$ [panel (b)] versus the spin polarization degree $p$. The other parameters are $\Delta\mu_p=10meV$, $\Delta\mu_s=2meV$, $\epsilon_\uparrow=0.1eV$, $\epsilon_\downarrow=0.2eV$, $\Gamma_0=10meV$, $T_L=300K$, and $T_R=500K$. Panel (c) shows the charge and spin efficiency $\bar{\eta}_{p/s}$ and panel (d) depicts the charge and spin power $\langle \dot{w}_{p/s}\rangle$ versus the quantum dot energy level $\epsilon$. The onsite spin-dependent energy $\epsilon_\sigma = \epsilon +\frac{1}{2}\sigma\Delta\epsilon$ with $\Delta\epsilon=0.06eV$. The other parameters are $\Delta\mu_p=10meV$, $\Delta\mu_s=2meV$, $p=0.2$, $\Gamma_0=10meV$, $T_L=300K$, and $T_R=500K$.}
    \label{3eta}
\end{figure}

In Fig.~\ref{3eta}(a) and (b), we plot the charge and spin current efficiency along with the corresponding power versus the spin polarization degree $p$. We observe that the engine enters the dud regime around polarization $p=-1$. In this region the transmission coefficient for spin-up electrons $\mathcal{T}_\uparrow$ [Eq.~\eqref{eq:13}] is negligible (since $\Gamma_{R\uparrow} = \Gamma_0 (1+p)\approx 0$) and hence only the spin-down current contributes to the charge and spin current. This causes the spin-current to be negative even though the spin bias is positive leading to a negative spin power and hence a negative spin efficiency $\bar{\eta}_s$. This precisely is the condition for the engine to become dud even though the charge efficiency $\bar{\eta}_p$ is well defined.



Next we vary the parameter $\epsilon$ [Fig.~\ref{3eta}(c) and (d)] that changes the spin-dependent energy $\epsilon_{\uparrow}\in [0.06, 0.1]eV$ and $\epsilon_{\downarrow} \in [0,0.04]eV$ of the QDSV such that $\epsilon_\uparrow > \epsilon_\downarrow$ due to a positive energy level splitting $\Delta\epsilon$. The spin currents will be chemical potential driven from left to right if $\mu_{L\sigma} > \epsilon_{\sigma} > \mu_{R\sigma} = 0eV$ due to the resonant or elastic tunneling of spins. The spin-up chemical potential of the left lead ($\mu_{L\uparrow} = 0.012 eV$) is such that it is never greater than the spin-up energy $\epsilon_{\uparrow}$ causing the up-current to be entropy driven from right to left due to the temperature bias $T_R > T_L$. On the other hand, the variation in $\epsilon$ is such that within the range $[0.03, 0.038)eV$ we have resonant tunneling of down spins from left to right since $\mu_{L\downarrow} > \epsilon_{\downarrow} > \mu_{R\downarrow}$. In this range of $\epsilon$ since the spin-up and -down currents are opposite in direction the charge current $I_c=I_{\uparrow}+I_{\downarrow}$ decreases in magnitude whereas the spin current $I_s=I_{\uparrow}-I_{\downarrow}$ increases. This obviously causes the spin power to dominate the charge power within this range of $\epsilon \in [0.03, 0.038)eV$ with the crossover being at $\epsilon = 0.038 eV$ beyond which charge power dominates. It is important to note here that when spin power dominates the charge power the down-spin power ($\langle \dot{w}_{\downarrow}\rangle = -I_{\downarrow}\Delta\mu_{\downarrow}$) is negative. Hence, in this regime the machine to convert heat to spin polarized power is a dud engine, whereas the machine to convert heat to spin and charge power works as a thermoelectric engine.

\begin{figure}
  \includegraphics[width=\columnwidth]{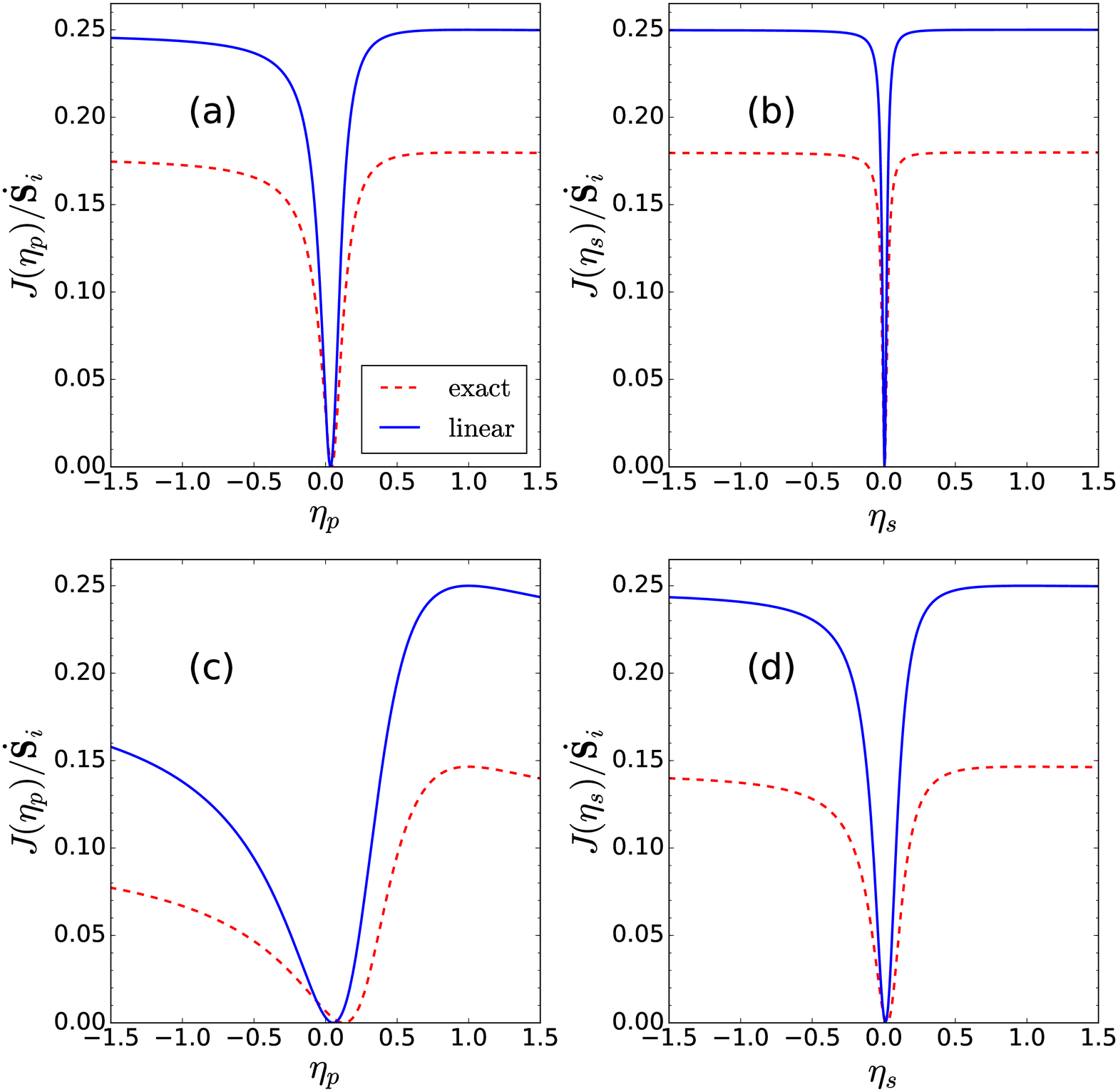}
  \caption{(Color online) Exact (red dashed line) and linear response (blue solid line) scaled efficiency LDFs $J(\eta)/\dot{\bf S}_i$ for spin [panels (b) and (d)] and charge [panels (a) and (c)] at different charge biases. Panels (a) and (b) correspond to $T_L=300K$, $T_R=500K$ and the charge bias $\Delta\mu_p=5meV$, and in panels (c) and (d) $T_L=300K$, $T_R=600K$ and $\Delta\mu_p=40meV$. For all panels the spin bias is chosen to be $\Delta\mu_s=0.2\Delta\mu_p$ and the other parameters are $\epsilon_\uparrow=0.3eV$, $\epsilon_\downarrow=0.4eV$, $p=0.2$, $\Gamma_0=10meV$.}
    \label{4EF}
\end{figure}

In Fig.~\ref{4EF}, we display the LDF of charge and spin efficiency which are scaled by the system EPR $\dot{\bf S}_i$ with the affinity values chosen close to the linear response limit in panels (a) and (b), and far beyond the linear response limit in panels (c) and (d). The efficiencies corresponding to the minimum values of the scaled LDF are the macroscopic ones, which are the most likely values in the statistics. It is most unlikely to achieve Carnot efficiency for either the spin or charge, since the maximum of the scaled LDF occurs always at the Carnot.

Similar to the macroscopic efficiencies wherein the charge efficiency $\bar{\eta}_p$ is always larger than the spin efficiency $\bar{\eta}_s$, the statistics shows that a broader range of charge efficiencies are likelier than their spin counterpart. The magnitudes of charge and spin efficiency fluctuation, as seen by the broadening of the scaled LDF, increases with the bias. This is mainly because an increase in bias leads to larger fluctuations in the currents and hence broader scaled LDF as also observed in traditional thermoelectric setups that do not posses the spin degree of freedom \cite{EF1, EF2, EF3, Bijay}. In the linear response limit, both the scaled efficiency LDFs, $J(\eta_{p(s)})/\dot{\bf S}_i$, take the upper-bound value of $1/4$ at the Carnot $\eta_{p(s)} = 1$, an analytic result discussed in Sec.~\ref{sec:IV}. However, the exact result of both charge and spin LDFs are always smaller than $1/4$. Moreover, since the fluctuations would have a stronger effect far from equilibrium we always find the linear response approximation underestimates the broadening of both the LDFs.

\section{Conclusion}
\label{sec:VI}
In this work, using the nonequilibrium Green's function approach, we obtained the scaled cumulant generating function (SCGF) of the thermoelectric quantum-dot spin valve in the collinear configuration. The Green's function approach allowed us to obtain the currents exactly in the QDSV setup and the SCGF provided a basis to obtain efficiencies and their statistics. In the linear response regime, using the conservation laws \cite{Matteo1} we obtained the correct Onsager matrix \cite{SSE2, Bauer2} and the reciprocal relations were a result of the underlying fluctuation theorem. We then provided a thermodynamic basis to define charge and spin current efficiencies from the EPR perspective. This not only allowed us to obtain the macroscopic efficiencies $\bar{\eta}_{p/s}$, but also justified our approach for the efficiency statistics via the large deviation function. 

In the linear response regime we were further able to extend the thermoelectric notion of FOM to the QDSV setup, wherein we require a spin $Z_sT$ and charge $Z_p T$ FOM. Interestingly, the maximum efficiencies were linked to the FOM\emph{s} via the same mathematical form as that in the traditional thermoelectric setup, implying that an infinite spin and charge FOM would help achieve the Carnot efficiency for spin and charge. In this regime, we also obtained the efficiency at maximum power that could be connected to the elements of the Onsager matrix.

Furthermore, we employed numerical techniques to compare between the exact methods and their linear response counterparts. We found that the linear response regime always underestimates the macroscopic efficiencies, average power, and also the efficiencies at maximum power. In regimes outside the linear response it even leads to wrong predictions and could lead to ill-defined results like a dud engine being highly efficient. Thus, the device performance can be accurately gauged only by the exact method and moreover the device can be made more efficient only far from equilibrium. Lastly, we looked at the efficiency large deviation functions wherein the macroscopic efficiencies corresponded to the minimum value of the LDFs. The most unlikely efficiency always turned out to be the Carnot efficiency. The scaled efficiency LDFs are bounded by the value $1/4$ taken at $\eta_{p(s)} =1$ in the linear response regime.

Overall, the method outlined in this work provides a strong basis to explore a QDSV connected to electrodes with a spin-dependent temperature (see appendix for details) or when the left electrode is not parallel or anti-parallel to the right electrode (non-collinear setup). In such a non-collinear setup, the spin current can transfer spin angular momentum and induce a spin transfer torque which could be used to switch the magnetic orientation of the ferromagnetic layers \cite{Theodonis1, Theodonis2, Theodonis3}. This phenomenon is a result of spin-flip processes that leads to spin currents with $x$, $y$, and $z$ polarizations. Similar to the collinear setup discussed in this work the $z$ polarized current would still be conserved, but the unconserved $x$ and $y$ polarized currents would induce a spin torque. Depending on the polarization, a parallel ($\tau_{\parallel}$) and perpendicular ($\tau_\perp$) torque would act on the system-electrode interface in the QDSV. These additional torques are the extra affinities that would cause the Onsager matrix to be a $5\times 5$ matrix. Such a non-collinear system would then form a perfect test bed to explore the fluctuation theorem for a complicated setup and gain insight as to how the Onsager matrix would transform from a $5\times 5$ matrix to a $3\times 3$ matrix when the angle of polarization between the electrodes is varied. Moreover, our discussion was limited to the electronic contributions to thermospin transport and in order to connect to real devices a promising future avenue would be to tackle the effects of electron-phonon interaction.

\begin{acknowledgements}
G.-T. and J.-W. are financially supported by NSF-China (Grant No. 11374246), the GRF (Grant No. 17311116), and the UGC (Contract No. AoE/P-04/08) of the Government of HKSAR. J.-T. is supported by the European Research Council project NanoThermo (ERC-2015-CoG Agreement No. 681456). The authors would like to thank Xiaobin Chen for useful discussions and Massimiliano Esposito for valuable feedback on the manuscript.
\end{acknowledgements}

\section*{APPENDIX: spin-dependent voltage and temperature bias}
In the appendix, we generalize the formalism in the main text to the case with both spin-dependent voltage bias and temperature gradient, $\Delta T_s$. Spin dependent temperature $T_{\alpha\sigma}$ in $\alpha$-electrode is due to spin polarized heat current \cite{Ts}.
Because of the spin-dependent temperature gradient, we shall have two separate counting fields $\lambda_{E\uparrow}$ and $\lambda_{E\downarrow}$ to count the energy current carried by the spin up and down electrons in the left electrode. Analogously to Eq.~\eqref{eq:12}, we have
\begin{widetext}
\begin{equation}
\label{eq:57}
{\cal F}(\lambda_\uparrow, \lambda_\downarrow, \lambda_{E\uparrow}, \lambda_{E\downarrow})
= \int \frac{d\omega}{2\pi} \ln \Big\{ 1 + \sum_\sigma \left[
(e^{ i\lambda_\sigma +i\omega\lambda_{E\sigma}}-1) {\cal T}_\sigma f_{L\sigma}(1-f_{R\sigma}) +
(e^{-i\lambda_\sigma -i\omega\lambda_{E\sigma}}-1) {\cal T}_\sigma f_{R\sigma}(1-f_{L\sigma})\right] \Big\} .
\end{equation}
\end{widetext}
One could verify the fluctuation theorem with the symmetry,
\begin{align}
\label{eq:58}
&{\cal F}(\lambda_\sigma,\lambda_{E\sigma}) = \notag \\
&{\cal F}(-\lambda_\sigma+i(\beta_R\mu_{R\sigma}-\beta_L\mu_{L\sigma}),-\lambda_{E\sigma}+i(\beta_{L\sigma}-\beta_{R\sigma}) ).
\end{align}
From Eq.~\eqref{eq:57}, one can get the spin-$\sigma$ current $I_{\sigma}$ and heat current $I_{h\sigma}$ in the linear regime as following,
\begin{equation} \label{AA1}
\begin{bmatrix}
&I_{\sigma}  \\  &I_{h \sigma}
\end{bmatrix} =
\mathbf{L}_\sigma
\begin{bmatrix}
&\Delta\mu_\sigma/T  \\  & -\Delta T_\sigma/T^2
\end{bmatrix} ,
\end{equation}
with $\Delta T_\sigma = T_{R\sigma}-T_{L\sigma}$.
The spin-$\sigma$ Onsager matrix $\mathbf{L}_\sigma$ is given by
\begin{equation}
\mathbf{L}_\sigma =
\begin{bmatrix}
L_\sigma & L_{\sigma h}  \\
L_{\sigma h} & L_{\sigma hh}
\end{bmatrix}
\end{equation}
with matrix entries
\begin{align}
L_{\sigma} &= -\int \frac{d\omega}{2\pi} f' {\cal T}_{\sigma} , \nonumber \\ 
L_{\sigma h}&=-\int\frac{d\omega}{2\pi} (\omega-\mu_{R\sigma}) f'{\cal T}_{\sigma},\\
L_{\sigma hh} &= -\int\frac{d\omega}{2\pi} (\omega-\mu_{R\sigma})^2 f' {\cal T}_{\sigma} .\nonumber
\end{align}
The Onsager coefficients are related to the transport coefficients,
\begin{align}
G_\sigma &= \left.\frac{I_{\sigma}}{\Delta\mu_{\sigma}}\right|_{\substack{\Delta T_\sigma = 0}} = \frac{L_{\sigma}}{T}\\
K_\sigma &= \left.-\frac{I_{h\sigma}}{\Delta T_\sigma}\right|_{I_\sigma = 0} = \frac{1}{T^2}\frac{\mathrm{det} \,\mathbf{L}_\sigma}{L_\sigma} \\
S_\sigma &= \left.\frac{\Delta \mu_\sigma}{\Delta T_\sigma}\right|_{I_{\sigma} = 0} = \frac{1}{T}\frac{L_{\sigma h}}{L_{\sigma}}, 
\end{align}
where $G_{\sigma}$ is the spin-$\sigma$ conductance, $K_\sigma$ is the spin-$\sigma$ thermal conductance, and $S_\sigma$ is the spin-$\sigma$ Seebeck coefficient.

In order to obtain the SCGF for the charge power, spin power, spin-up heat current, and spin-down heat current, we make the following substitutions
\begin{align}
\label{eq:59}
i\lambda_\uparrow &\rightarrow -\lambda_{h\uparrow}\mu_{R\uparrow}-\lambda_p\Delta\mu_p -\lambda_s\Delta\mu_s ,  \notag \\
i\lambda_\downarrow &\rightarrow -\lambda_{h\downarrow}\mu_{R\downarrow}-\lambda_p\Delta\mu_p +\lambda_s\Delta\mu_s , \notag \\
i\lambda_{E\sigma} &\rightarrow \lambda_{h\sigma} ,
\end{align}
that are similar to Eq.~\eqref{eq:44}. Further assuming $\beta_{L\uparrow} = \beta_{L\downarrow}=\beta_L$, so that the spin-dependent temperatures only exist in the hot reservoir, we have the symmetry relation for SCGF,
\begin{align}
\label{eq:60}
&{\cal F}(\lambda_p,\lambda_s,\lambda_{h\uparrow},\lambda_{h\downarrow}) =   \notag \\
&{\cal F}(-\lambda_p+\beta_L,-\lambda_s+\beta_L,-\lambda_{h\uparrow}-\Delta\beta_\uparrow,-\lambda_{h\downarrow}-\Delta\beta_\downarrow) .
\end{align}
with $\Delta\beta_\sigma = \beta_{L} -\beta_{R\sigma}$. Due to the presence of an additional independent affinity, i.e., spin-dependent temperature gradient, the Onsager matrix becomes a $4 \times 4$ matrix \cite{thermo_SC} given by
\begin{equation}
\begin{bmatrix}
&I_p  \\ & I_{h\uparrow} \\ &I_s  \\ & I_{h\downarrow}
\end{bmatrix} =
\begin{bmatrix}
G T & P\alpha T & PG T & \alpha T \\
L_{\uparrow h} & L_{\uparrow hh} & L_{\uparrow h} & L_{\uparrow hh} \\
PG T & \alpha T & G T & P\alpha T \\
L_{\downarrow h} & L_{\downarrow hh} & -L_{\downarrow h} & -L_{\downarrow hh}
\end{bmatrix}
\begin{bmatrix}
&\Delta\mu_p/T \\  & -\Delta T_{\uparrow}/T^2 \\  & \Delta\mu_s/T \\ & -\Delta T_{\downarrow} /T^2
\end{bmatrix} ,
\end{equation}
with $\Delta T_{\sigma} = T_{R\sigma} - T_{L\sigma}$ ($\sigma = \uparrow, \downarrow$), $\Delta\mu_{(p,s)}$ defined in Eq.~\eqref{eq:22}, and $T=\sum_\sigma (T_{R\sigma} + T_{L\sigma})/4$. Performing a transformation to the spin-dependent heat currents,
\begin{align}
I_{h}&=I_{h\uparrow}+I_{h\downarrow} \nonumber \\
I_{hs}&=I_{h\uparrow}-I_{h\downarrow}
\end{align}
we obtain
\begin{equation}
\label{eq:61}
\begin{bmatrix}
&I_p  \\ & I_{h} \\ &I_s  \\ & I_{hs}
\end{bmatrix} =
\begin{bmatrix}
G T & P\alpha T & PG T & \alpha T \\
P\alpha T & G_h T & \alpha T & PG_h T \\
PG T & \alpha T & G T & P\alpha T \\
\alpha T & PG_h T & P\alpha T & G_h T
\end{bmatrix}
\begin{bmatrix}
&\Delta\mu_p/T \\  & -\Delta T/T^2 \\  & \Delta\mu_s/T \\ & -\Delta T_s /T^2
\end{bmatrix}.
\end{equation}
The coefficients above are expressed by
\begin{align*}
G &= \sum_\sigma\frac{L_{\sigma}}{T}, \quad PG = \sum_\sigma\frac{\sigma L_{\sigma}}{T} \\
\alpha &= \sum_\sigma\frac{\sigma L_{\sigma h}}{T} , \quad P\alpha = \sum\frac{L_{\sigma h}}{T}  \\
G_h &= \sum_\sigma\frac{L_{\sigma hh}}{T} , \quad PG_h = \sum\frac{\sigma L_{\sigma hh}}{T}
\end{align*}
which are obtained from Eqs.~\eqref{AA1} and with $\sigma = +1$ for $\uparrow$ and $\sigma=-1$ for $\downarrow$.
Above the temperature gradients are given by
\begin{equation}
\label{eq:62}
\Delta T= \sum_\sigma\frac{T_{R\sigma}-T_{L\sigma}}{2} , \quad
\Delta T_s= \sum_\sigma \frac{\sigma(T_{R\sigma}-T_{L\sigma})}{2}.
\end{equation}

The entropy production rate of the system in the steady state is expressed as,
\begin{equation}
\label{eq:63}
\dot{\bf S}_i= \sum_{\alpha ,\sigma}\beta_{\alpha\sigma} I_{\alpha h\sigma} \geq 0,
\end{equation}
with the spin-$\sigma$ heat current in the $\alpha$th electronic reservoir $I_{\alpha h\sigma}=I_{\alpha E\sigma}-\mu_{\alpha\sigma} I_{\alpha\sigma}$. Having assumed $\beta_{L\uparrow} = \beta_{L\downarrow}=\beta_L$ and using the particle, spin, spin-up heat, and spin-down heat current conservation laws $I_{L (\cdot)}+I_{R (\cdot)}=0$ with $(\cdot) = s,p,h\uparrow, h\downarrow$, we can rewrite the EPR as,
\begin{align}
\label{eq:64}
\dot{\bf S}_i &= \dot{\bf S}_{h\uparrow} +\dot{\bf S}_{h\downarrow} + \dot{\bf S}_p + \dot{\bf S}_s \notag \\
\dot{\bf S}_{h\sigma}&=-(\beta_L-\beta_{R\sigma}) I_{h\sigma} \notag \\
\dot{\bf S}_{p(s)}&=\beta_L I_{p(s)} \Delta\mu_{p(s)},
\end{align}
with $I_{h\sigma} = I_{E\sigma}-\mu_{R\sigma} I_\sigma$. Spin-up and -down currents without electrode indexes refer to the currents in the left electrode. In order to obtain the above partition of the EPR with the form like Eq.~\eqref{eq:29} from Eq.~\eqref{eq:63}, spin-dependent temperature should be assumed to exist only in the hot reservoir. Hence, we can define the charge current efficiency $\bar{\eta}_p$ and spin current efficiency $\bar{\eta}_s$ in the QDSV as
\begin{align}
\label{eq:65}
0\leq \bar{\eta}_p &= \frac{ -\dot{\bf S}_p }{ \dot{\bf S}_{h\uparrow} + \dot{\bf S}_{h\downarrow} + \dot{\bf S}_s } = \frac{-\langle \dot{w}_p\rangle}{\sum_\sigma \eta_{C\sigma} I_{h\sigma} +\langle \dot{w}_s\rangle} \leq 1,\notag\\
0\leq \bar{\eta}_s &= \frac{ -\dot{\bf S}_s }{ \dot{\bf S}_{h\uparrow} + \dot{\bf S}_{h\downarrow} + \dot{\bf S}_p }=\frac{-\langle \dot{w}_s\rangle}{\sum_\sigma \eta_{C\sigma} I_{h\sigma} +\langle \dot{w}_p\rangle} \leq 1,
\end{align}
where the charge and spin power are given by 
\begin{align}
\langle \dot{w}_p\rangle &= -I_p\Delta\mu_p, \quad \langle \dot{w}_s\rangle = -I_s\Delta\mu_s
\end{align}
with the Carnot efficiency of spin direction $\sigma$ defined as $\eta_{C\sigma} = 1- T_L/T_{R\sigma}$. Using the procedure outlined in Sec.~\ref{sec:IV}, the charge and spin efficiency LDF could also be obtained from the SCGF, for the case of spin-dependent temperature gradient.

\end{document}